\documentclass[%
 reprint,
 amsmath,amssymb,
 aps,
]{revtex4-2}
\usepackage{amsmath}
\usepackage{graphicx} 
\usepackage{amssymb}
\usepackage{lipsum}
\usepackage{float}
\usepackage{amsmath}

\usepackage[markup=default]{changes}
\definechangesauthor[name=Igor Lobanov, color=orange]{LIS}

\usepackage[unicode=true,colorlinks=true,citecolor=blue,urlcolor=blue]{hyperref}

\usepackage{bm}
\usepackage{color}
\usepackage{epsfig}

\usepackage[normalem]{ulem}

\renewcommand {\phi}{{\varphi}}

\begin{document}
\title{Skyrmion dynamics in moir{\'e} magnets}

\author{P.~S.~Shaban}
\affiliation{Department of Physics, ITMO~University, St.~Petersburg, 197101, Russia}
\author{I.~S.~Lobanov}
\affiliation{Department of Physics, ITMO~University, St.~Petersburg, 197101, Russia}
\author{V.~M.~Uzdin}
\affiliation{Department of Physics, ITMO~University, St.~Petersburg, 197101, Russia}
\author{I.~V.~Iorsh}
\email{i.iorsh@metalab.ifmo.ru}
\affiliation{Abrikosov Center for Theoretical Physics, MIPT, Dolgoprudnyi, Moscow Region 141701, Russia}
\affiliation{Department of Physics, ITMO~University, St.~Petersburg, 197101, Russia}

\begin{abstract}
We consider a twisted magnetic bilayer subject to the perpendicular electric field. 
The interplay of induced Dzyaloshinskii - Moriya interaction and spatially varying moir{\'e} exchange potential 
results in complex non-collinear magnetic phases in these structures. 
We numerically demonstrate the coexistence of intralayer skyrmions and bound interlayer skyrmion pairs 
and show that they are characterized by distinct dynamics under the action of external in-plane electric field. 
Specifically we demonstrate the railing behaviour of skyrmions along the domain walls which could find 
applications in spintronic devices based on van der Waals magnets.
\end{abstract}

\maketitle 

\section{I. Introduction}
Van der Waals (vdW) materials offer unprecedented opportunities to form heterostructures of different monolayers with unique magnetic, transport, and optical properties and enable a powerful toolbox for the bottom-up material engineering~\cite{liu2016van,novoselov20162d}. 

VdW magnets are a relatively novel class of the vdW materials~\cite{burch2018magnetism,blei2021synthesis}. The first experimental realization of two-dimensional vdW magnets, CrI$_3$~\cite{huang2017layer} and CrGeTe$_3$~\cite{gong2017discovery} was reported in 2017 and since then the family of 2D magnets is rapidly growing with dozens of new materails appearing each year~\cite{yang2021van}. Due to the atomic-scale thickness, vdW magnets are highly susceptible to the external perturbations such as external fields~\cite{jiang2018electric,polshyn2020electrical,jiang2018controlling} and strain~\cite{li2019pressure,qi2023recent}. Specifically, perpendicular electric field may induce the Dzyaloshinskii - Moriya interaction (DMI) which leads to the emergence of non-collinear magnetic structures~\cite{jaeschke2021theory} such as helices, individual skyrmions and skyrmion crystals~\cite{behera2019magnetic}.   Moreover,  for some vdW materials, their magnetic properties depend crucially on stacking configuration: controlling the stacking angle and relative displacement of individual magnetic monolayers allows for the precise tuning of the interlayer exchange coupling as well as dipole-dipole interaction which results in the emergence of various new magnetic phases in twisted vdW magnets~\cite{chen2019direct,sivadas2018stacking, tong2018skyrmions,huang2020emergent,song2021direct,xu2020topological,xiao2021magnetization}. 

In twisted magnetic bilayers, the site-dependent interlayer exchange potential, or moir{\'e} potential, which is periodic with a period equal to the moir{\'e}  supercell, defines the spatial scale of the emergent non-collinear phases. At the same time, the chiral interactions such as DMI correspond to an  alternative spatial scale which can be tuned by external perpendicular electric fields. It is thus tempting to explore the emergent magnetic phases in twisted magnetic bilayers where both interlayer exchange moir{\'e} potential and intralayer DMI are present. Such a competition would lead to a rich phase diagram of such structures, 
if the strength and characteristic lengthscale of two types of interactions are compatible.
\begin{figure}[!h]
\centering
\includegraphics[width=0.49\textwidth]{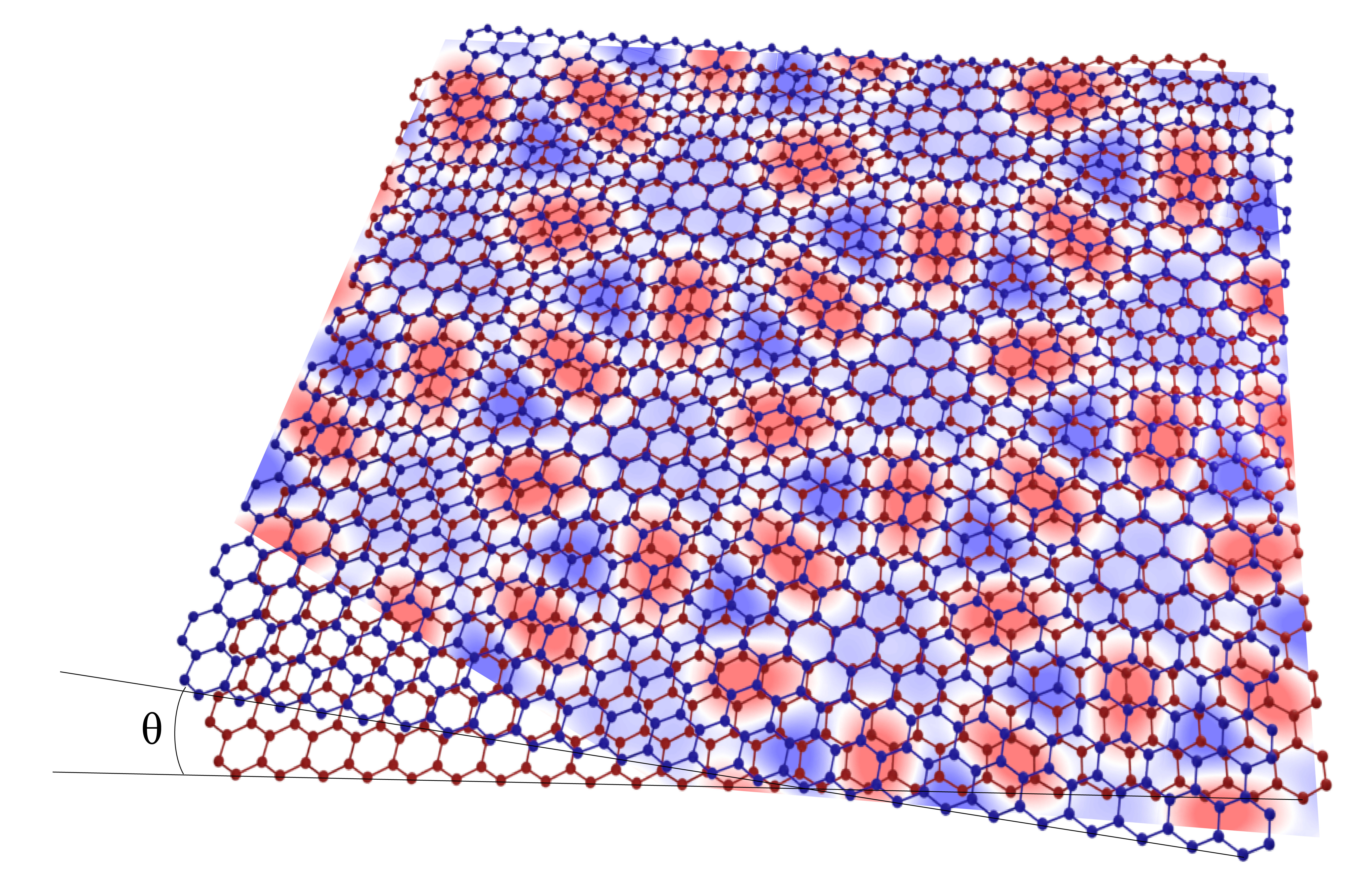}
\caption{\label{fig:Fig1}Twisted ferromagnetic bilayer. 2D color map demonstrates interlayer interaction potential (moir{\'e} potential). Moir{\'e} period is proportional to $a/\theta$, $a$ is the lattice period, $\theta$ is the angle between twisted layers. Twist angle on the picture is much bigger than the real moir{\'e} angle in simulations and used for better visual perception.}
\end{figure}
A similar competition of intra- and interlayer interactions takes place in synthetic antiferromagnets \cite{Duine18}.
In these structures, two magnetic layers are exchange-coupled through a thin metal non-magnetic spacer due to the Ruderman-Kittel-Kasuya-Yosida (RKKY) interaction of conduction electrons. The magnitude and sign of the interlayer exchange coupling (IEC) depend on the thickness of the nonmagnetic interlayer, which can be chosen 
in such a way as to ensure antiferromagnetic (AF) exchange. If in one of the magnetic layers there are domains with different directions of magnetization, then 
in the other layer the same domain structure will be repeated, but with the opposite direction of magnetization.

The IEC varies depending on the thickness of the spacer layer. If it is made in the form of a wedge, then the IEC will oscillate with thickness, which will lead to the formation of a domain structure, even if one of the magnetic films is in a single-domain state. Such a smooth change in thickness makes it possible to observe short-range and long-range IEC  oscillations in metal magnetic trilayers \cite{Unguris91}.
If, in the presence of DMI, localized magnetic structures of the skyrmion type are formed in one magnetic layer, then another skyrmion should form in the second layer, which will be in a bound state with the first one \cite{Legrand19}. The skyrmions will be coupled ferromagnetically or antiferromagnetically, and this bonding should also vary with the thickness of the nonmagnetic interlayer.
The same behavior should be observed in twisted vdW magnetic layers. This article will present the features of the magnetic structure and dynamics of skyrmions in a moire magnet associated with the competition of in-plane and out-of-plane interactions.
We consider a structure shown in Fig.~\ref{fig:Fig1}. A twisted magnetic bilayer is subject to an external electric field inducing intralayer DMI in each layer. The interlayer moir{\'e} potential arises due to the spatially dependent IEC. The system models the recently reported experiments with twisted monolayers of $\mathrm{CrI_3}$~\cite{xu2022coexisting,wang2022magnetic}. We start from numerical modelling of the magnetic phases supported by this structure. We show that spatially dependent interlayer exchange potential results in complex non-collinear magnetic structures and the formation of Ferromagnetic (FM) and antiferromagnetic domains (AF). Moreover, we show that the system supports several types of intralayer skyrmions as well as bound interlayer skyrmion pairs predicted previously~\cite{tong2018skyrmions}. We then study the skyrmion dynamics using the Landau-Lifsthis-Gilbert (LLG) equation and demonstrate  and demonstrate the railing of skyrmions along the domain walls under the action of external in-plane electric field. We give a qualitative explanation of the observed effect using the Thiele equation.

The article is organised as follows: in Section II we define the model and present the results of numerical simulations of the magnetic phases supported by the structure. In Section III we present the results on the dynamics of the skyrmions under the external in-plane electric field. Section IV summarizes the obtained  results.
\section{II. Non collinear magnetic phases in twisted magnetic bilayer}
\subsection{Model}

In our model, we consider two layers of a ferromagnetic material with a hexagonal crystal lattice, rotated relative to each other by a small angle, which determines the shape of the moir{\'e} potential. In the ferromagnetic case considered below, the interaction potential is rather difficult to describe analytically, but it can be specified numerically. 
Moir{\'e} period is proportional to $a/\theta$, $a$ is the lattice period, which is typically about several angstroms. 
The twist angle we use in our calculations equals to approximately 0.7$^{\circ}$. Profile of moir{\'e} potential $\Phi (\mathbf{r})$ 
adopted from ~\cite{hejazi20} is shown in Fig.~\ref{fig:Fig2}a along with scale bar.

We consider a continuous generalized Heisenberg-type model with the energy

\begin{multline}
    E = d\cdot\int d^2\mathbf{r} \Bigg[ \sum_{i = 1,2} \Big(\mathcal{A}(\nabla\mathbf{n}_i(\mathbf{r}))^2 - \mathcal{K}n_{zi}^2(\mathbf{r}) + \\ + \mathcal{D}\bm{n}_i(\bm{r})\cdot(\hat{\bm{z}}\times\nabla)\times\bm{n}_i(\bm{r}) \Big) - J_{1,2}\Phi (\bm{r}) \bm{n}_1(\bm{r})\cdot\bm{n}_2(\bm{r}) \Bigg]
\end{multline}

Here $\mathbf{n}_1$ and $\mathbf{n}_2$ are the unit vectors along the magnetization in layers 1 and 2, respectively, $d$ is the magnetic layer thickness. 
$\mathcal{A}$ is the exchange stiffness constant.
Pairs of nearest atoms in different layers also contribute to the Heisenberg exchange, 
but the interaction strength depends on the position of atoms and equals $J_{1,2}\Phi (\mathbf{r})$,
where $J_{1,2}$ is a parameter controlling interaction strength.

DMI is turned on in each layer,
but there is no antisymmetric exchange interaction between layers.
Dzyaloshinskii vectors are parallel to the line connecting interacting spins,
the length of the vector determines the DMI density $\mathcal{D}$.
Anisotropy axis $\mathbf e_z$ is the same for all points of the system, the anisotropy density 
$\mathcal{K}>0$ corresponds to the easy axis anisotropy. The spin texture generated in a moir{\'e} supercell can give rise to an electric polarization
associated with such a non-collinear magnetic state due to spin-orbit coupling,
resulting in a local ferroelectric order following moir{\'e}~\cite{otero2023moire}

When performing numerical calculations, the micromagnetic model is dicretized on a square lattice.
{A cell of $429\times50$ lattice points } with free boundary conditions was used. Its size coincided with the cell size 
in \cite{hejazi20}, where the moir{\'e} potential is taken from.

Micromagnetic parameters are converted into {the discrete model parameters }
\begin{equation}
    J=2 \mathcal{A},\quad D=a \mathcal{D},\quad K =a^2 \mathcal{K},
    \label{param}
\end{equation}
where $a$ is the in-plane lattice constant.

In our modeling, we use dimensionless variables, and all  parameters in (\ref{param}) are measured in  $J$-units. 
The easy axis anisotropy, $K/J =0.01$, is used below, which gives an estimate of approximately 22 lattice constants 
for the thickness of a domain wall in a bulk material without DMI: $L=\pi\sqrt {\mathcal{A}/\mathcal{K}}$.

DMI can be varied by changing the external electric field \cite{PhysRevB.103.174410}, so the system will 
be considered at different values of the DMI constant. In a bulk material with DMI, the ferromagnetic (FM) state becomes unstable 
with respect to the transition to the spiral state at $D_s=4\sqrt {\mathcal{A}\mathcal{K}}/\pi$. We will use the 
dimensionless parameter $\zeta=D/D_s$.

\subsection{Results}

\begin{figure}[H]
\center{\includegraphics[scale=0.26]{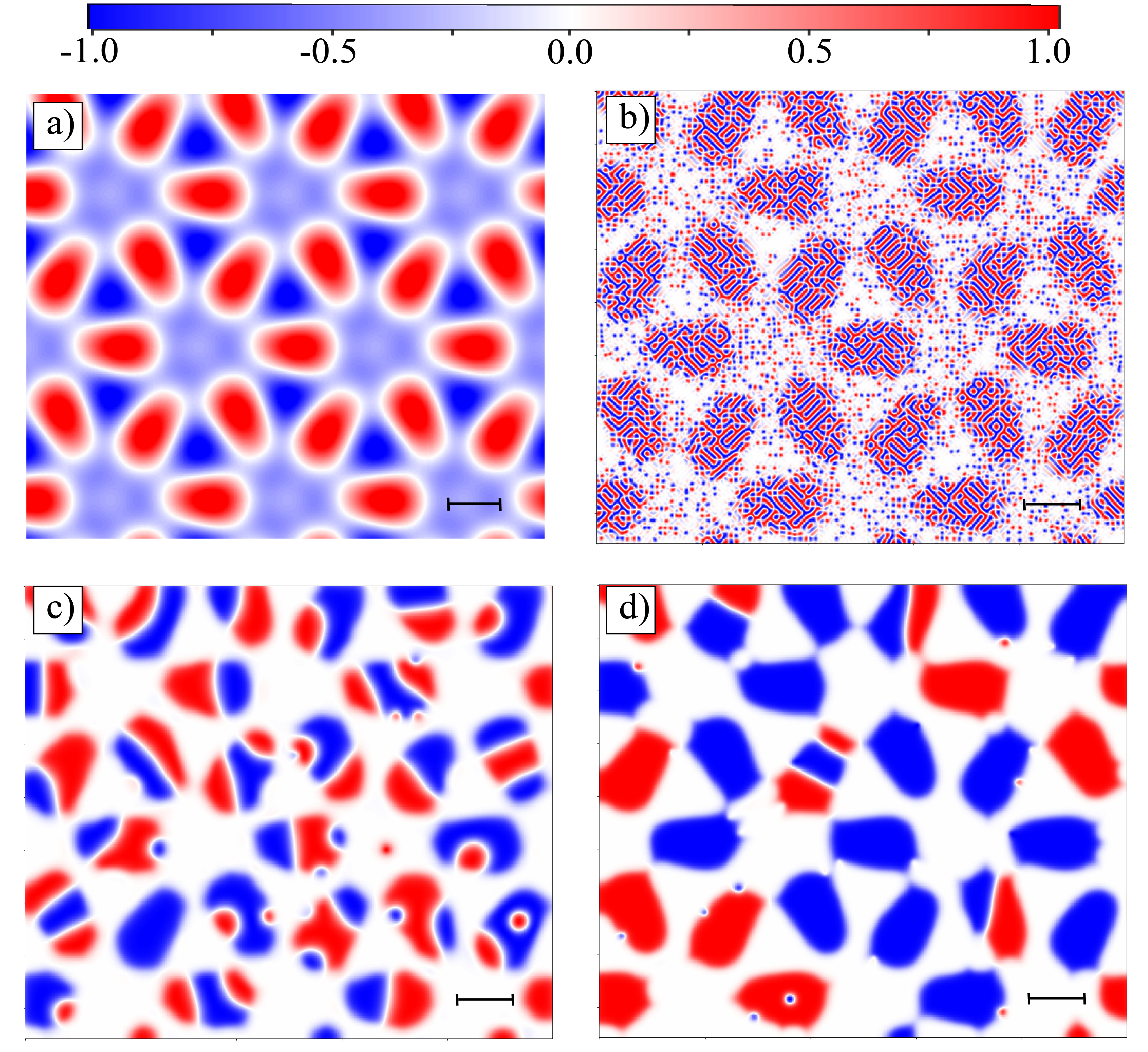}}
\caption{Spatial dependence of interlayer interaction $\Phi (\mathbf{r})$ (a) and normalized z-projection of total bilayer magnetic moments in moir{\'e} magnets for $\zeta=10$ (b), $\zeta=1.22$(c) and $\zeta=0.78$ (d). Scale bar: 100 lattice constants.}
\label{fig:Fig2}
\end{figure}

The determination of the magnetic configuration corresponding to the local energy minimum begins from a state with a random 
distribution of  magnetic moment directions. The non-linear conjugate gradient method is used
{for energy minimization with Hessian matrix evaluated in Cartesian coordinates \cite{Lobanov21CPC}}.
The minimization stops when the gradient becomes less than $10^{-5}$. 
The system has a large number of metastable states with close energies, 
and {Fig. }\ref{fig:Fig2} reproduces
the typical examples of locally stable magnetic configurations for different values of the $\zeta$ parameter.
Fig.~\ref{fig:Fig2}a shows the spatial configuration of the moir{\'e} potential, which defines the regions with FM and AF  IEC. 
The interlayer exchange potential was computed in~\cite{hejazi20}.
$\Phi (\mathbf{r})=1$ (red) and $\Phi (\mathbf{r})=-1$ (blue) correspond to the FM and AF exchange, respectively, and the 
white lines are the FM grain boundaries, where the exchange is close to zero.

For $\zeta=10$, the ground state in each layer is a spiral structure of the "fingerprints" type. In the FM IEC region, the same structure is observed for the total magnetization of both layers, as can be seen from Fig. \ref{fig:Fig2}b.
In the AF region, the magnetizations of the helical domains are largely compensated, except for the ends of the domains, where point out-of-plane magnetic states appear for total magnetization. These states, however, are not layer-localized structures, and their mobility is very limited, since their movement can only arise as a result of rearrangement of the helical structure in each layer. We provide a more detailed illustration of the formation of these point-like states arising due to the superposition of the domain walls in the two layers in the Supplemental Material.

As effective DMI stregnth$\zeta$ decreases, the size of the domains increases. For example, at  $\zeta=1.22$ shown in Fig.~\ref{fig:Fig2}(c), the domain size is comparable with the moir{\'e} grain size. Moreover, the skyrmion states are identified  in FM and AFM regions, and at the boundaries between FM and AFM.  

For $\zeta=0.78$, {the observed skyrmions have } size much smaller than 
the regions of a constant IEC sign. 
Fig.~\ref{fig:Fig2}d indicates
that most of these skyrmions are located at the boundary of the moir{\'e} grains, 
although they also can be found inside the grains.  The magnetization profiles for other values of $\zeta$ are shown in Supplemental Material illustrating gradual increase of the domain size with the decrease of $\zeta$.

\begin{figure}[H]
\center{\includegraphics[scale=0.5]{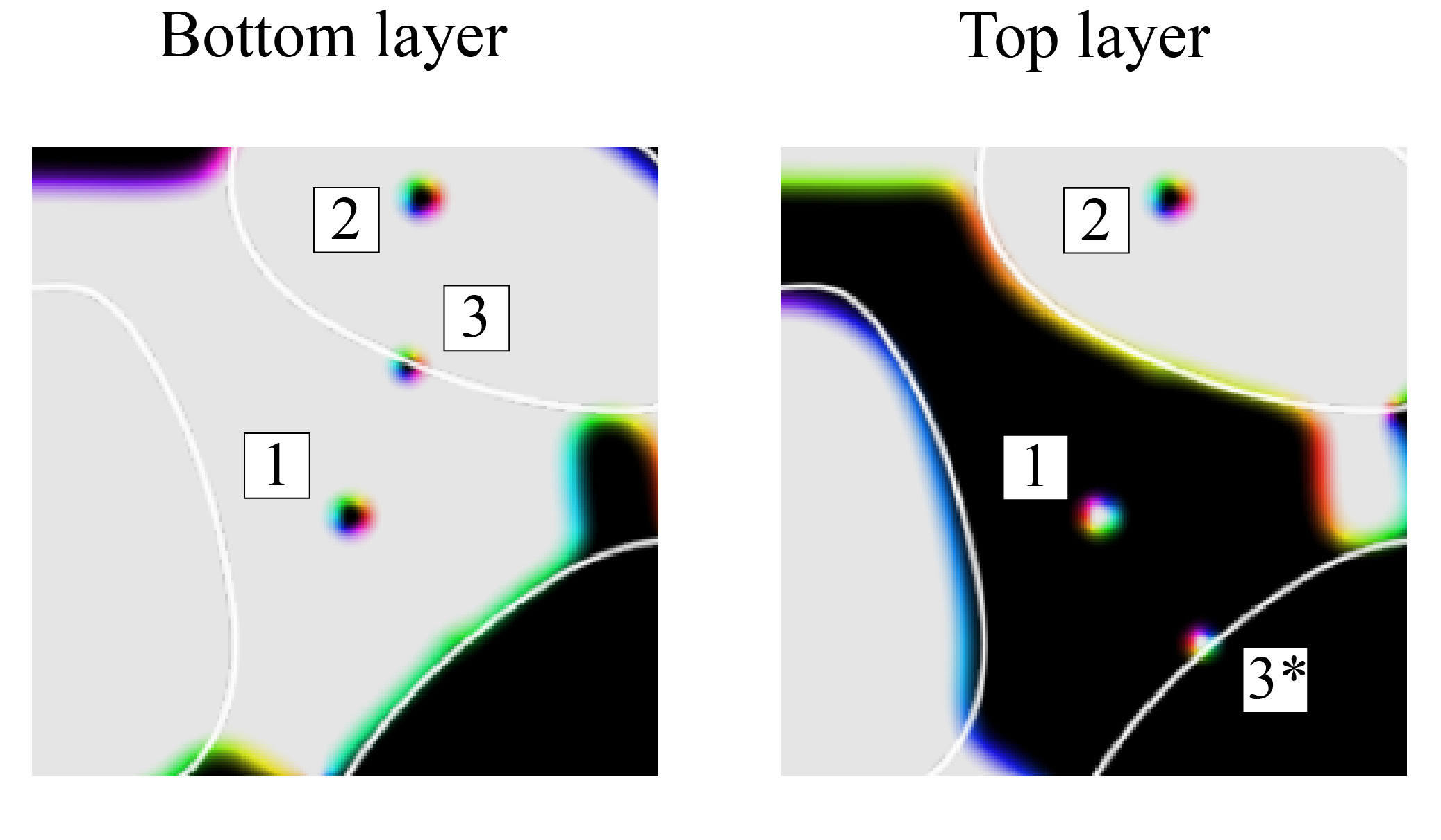}}
\begin{centering}
\caption{ Various types of skyrmions in a twisted bilayer. Light gray color corresponds to the direction of the magnetization vector out-of-plane and z-component is positive, black color is out-of-plane and negative z-component. Other colors demonstrate the orientation of the magnetization vector in-plane.
Pair of AF skyrmions located in the AF IEC grain (1), in the FM IEC grain (2) and single skyrmions (3,3*) 
fixed at the border of zero moir{\'e} potential in bottom and top layer respectively. 
Boundaries with zero moir{\'e} potential are shown as white lines.
}
\label{fig:Fig3}
\end{centering}
\end{figure}

In order to demonstrate different types of skyrmion structures, 
Fig. \ref{fig:Fig3} shows the configurations in the upper and lower layers of moir{\'e} magnet in this case.
Pairs of coupled skyrmions can reside both in the AF and FM IEC regions.
In the first case (1), the total topological charge of the pair is equal to zero, and in the second (2), to two.
Single skyrmions in one layer with a unit topological charge and domain walls in another layer are usually located in the region of zero moir{\'e} potential (3).

\begin{figure}[H]
\center{\includegraphics[scale=0.55]{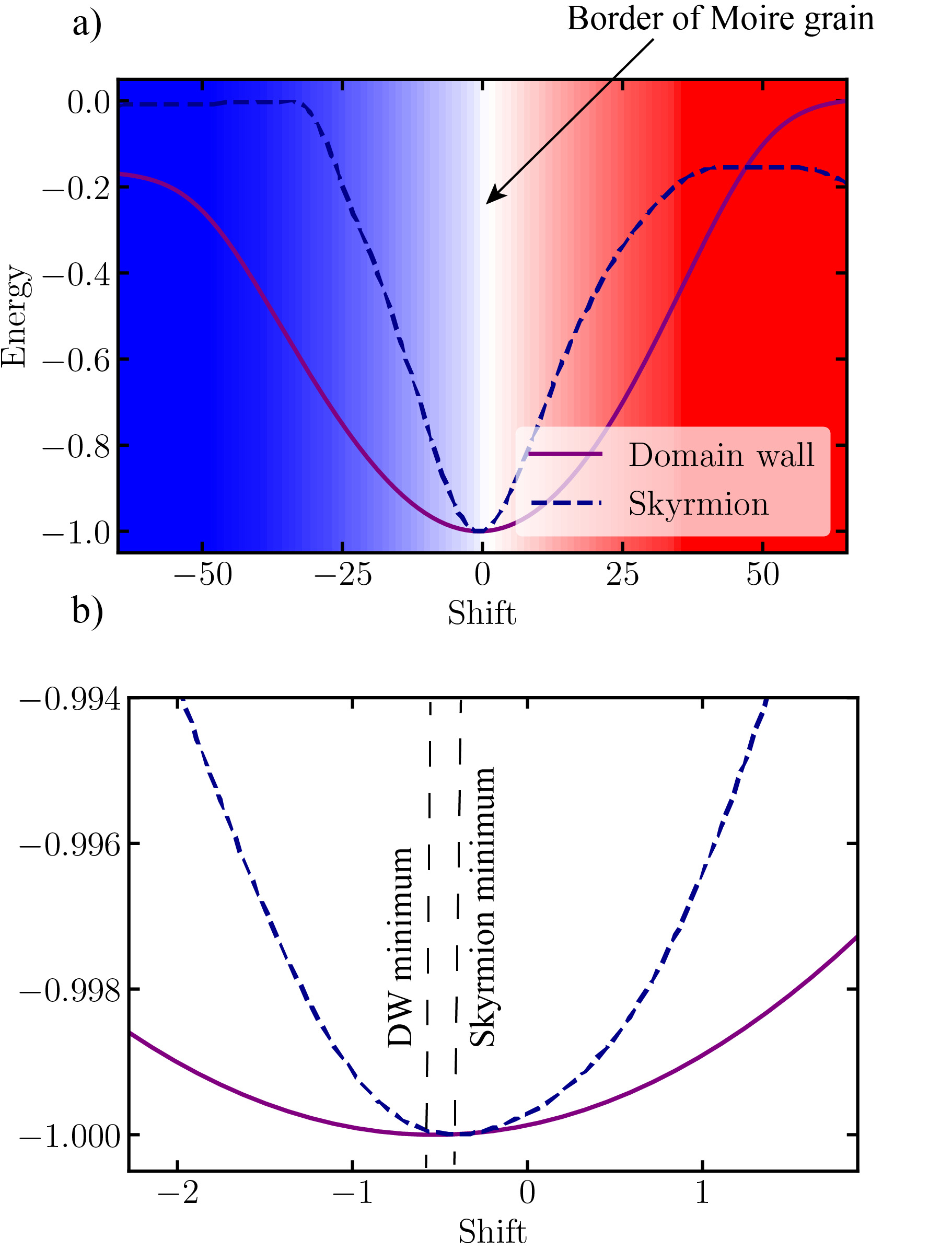}}
\begin{centering}
\caption{Dependence of domain wall and skyrmion energy on the displacement with respect to the interlayer exchange potential (IEC) boundary. (a) Blue and red shaded area correspond to the AFM and FM IEC. 
(b) {Zoomed central part presenting } the shift of skyrmion and domain wall potential  with respect to {the} grain boundary.}
\label{fig:Fig4}
\end{centering}
\end{figure}

To explain the localization of skyrmions and domain walls near the boundaries of moir{\'e} grains, the energies of these structures near the boundaries $\Phi (\mathbf{r})=0$ were calculated. The results are shown in Fig.~\ref{fig:Fig4}.
As a first step, we find the optimal position for the domain wall near zero moir{\'e} potential by varying its position along the line perpendicular to the grain boundary and calculating the energy of the system. It can be seen that at a certain position the energy is minimal, so this position is energetically favorable for the domain wall. It is also noticeable that this minimum is slightly offset from the point where $\Phi (\mathbf{r})=0$, which can be seen on the Figure \ref{fig:Fig4}.

The second step is to minimize the energy{as function of } the position of the skyrmion in the other layer. 
The wall in one layer remains {its position }, 
while the skyrmion in the other layer {is translated }
in the direction perpendicular to the moir{\'e} grain boundary. 
There is also a certain energy minimum here, which does not coincide with the minimum for the domain wall, 
but is located closer to zero of the moir{\'e} potential.

\section{III. Skyrmion dynamics in twisted magnetic bilayer}

In this section we will show that the three described types of skyrmions are characterized by the distinctive dynamics under applied electric field. We perform numerical simulation of the dynamics of the skyrmions using Landau-Lifshitz-Gilbert equation. At the same time, to provide a qualitative explanation of the specific features of the skyrmion dynamics, we employ the Thiele equation, which can be used when one can neglect the modification of the skyrmion internal structure. In the Supplemental Material we provide the results of the numerical simulation confirming that the this approximation is valid. Below we provide a brief derivation of the Thiele equation.
\subsection{Thiele equation}

The dynamics of the {bilayer} system is described by the Landau-Lifschitz-Gilbert (LLG) equation:
\begin{equation}
\frac{d \mathbf{n}}{dt}=-\gamma \mathbf{n}\times \left(\mathbf{H}_{eff}-\eta M_s \frac{d \mathbf{n}}{dt}\right)+\mathbf{\tau},
\end{equation}
where $\gamma$ is gyromagnetic ratio, $\eta$ is damping parameter{ and }
$\mathbf{H}_{eff}$ is the effective magnetic field
\begin{equation}
\mathbf{H}_{eff} = -\frac{1}{M_s}\frac{\partial E}{\partial \mathbf{n}}.
\end{equation}
All the vector fields depend on {the layer $l=1,2$} and {the spin coordinates} $\mathbf{r}=(x,y)$.
    The term $\mathbf{\tau}$ is Slonczewski spin transfer torque (STT)~\cite{PhysRevB.39.6995}:
\begin{equation}
    \mathbf{\tau} = -\mathbf{n}\times (\mathbf{n}\times \mathbf{j}_t)-\beta \mathbf{n}\times \mathbf{j}_t,
\end{equation}
where $\beta$ is anti-damping constant associated with STT and
\begin{equation}
\mathbf{j}_t = \bigg(\mathbf{j}\cdot \nabla\bigg)\mathbf{n}=\sum_{k=x,y}j_k\frac{\partial \mathbf{n}}{\partial k}.
\end{equation}

We are interested in the dynamics of the topological solitons assuming their shape is invariant.
Denote $\mathbf{R}^l=(R^l_x, R^l_y)$ the position of the soliton in the layer $l$.
If the shape is fixed, then $\mathbf{R}^l$ are the only varying degrees of freedom.
The constrained dynamics is derived by the projecting the velocity $\dot{\mathbf{n}}=d\mathbf{n}/dt$ 
to the generators of the translations of the solitons
\begin{equation}
\mathbf{G}^l_{k}=\frac{\partial \mathbf{n}_l}{\partial R^l_k}
=-\frac{\partial \mathbf{n}_l}{\partial k},\quad k=x,y.
\end{equation}

The projected LLG equation onto the space spanned by the vectors $G^l_k$ is called Thiele equation.
For the multilayer system the Thiele equation becomes:
\begin{equation}
\label{eq:thiele}
-4\pi Q^l J \dot{\mathbf{R}}^l = 
-\frac{\gamma}{M_s}\frac{\partial E}{\partial \mathbf{R}^l}
+\gamma\eta M_s A^l \dot{\mathbf{R}}^l
+4\pi Q^l  J \mathbf{j}^l+\beta^l A^l\mathbf{j}^l,
\end{equation}
where $Q^l$ is the topological charge of the layer $l$:
\begin{equation}
Q^l = \frac1{4\pi}\int \mathbf{n}_l\cdot\bigg(\frac{\partial \mathbf{n}_l}{\partial x}\times \frac{\partial \mathbf{n}_l}{\partial y}\bigg)dr,
\end{equation}
and we introduced matrices
\begin{align}
J=\begin{pmatrix}0 & 1 \\ -1 & 0 \end{pmatrix}
A^l_{jk} = \int \frac{\partial \mathbf{n}_l}{\partial j} \cdot \frac{\partial \mathbf{n}_l}{\partial k}dr
\quad j,k=x,y.
\end{align}

The Thiele equation can be solved with respect to $\dot{\mathbf{R}}$.
Consider FM pair of solitons assuming their perfect alignment,
then the system becomes essentially single layer with thicker layer.
Suppose the background phase is isotropic, e.g. FM, then $\partial E/\partial \mathbf{R}=0$.
The Thiele equation in this case is well-known:
\begin{equation}
\dot{\mathbf{R}} = -\big(4\pi Q J+\gamma\eta M_s A\big)^{-1}
\big(4\pi Q J+\beta A\big)\mathbf{j}.
\end{equation}
The soliton velocity in this case is connected with the current $\mathbf{j}$ by a linear transform,
and the transform commutes with rotations.
Therefore the angle between the soliton velocity $\dot{\mathbf{R}}$ and the current $\mathbf{j}$ is constant and is called Hall angle.

If the soliton is invariant under reflections with respect to both coordinate axes
(e.g. skyrmion, skyrmionium), then the matrix $A$ is proportional to the identity operator $\hat{I}$, $\hat{A}=A\hat{I}$.
The Hall angle is given by:
\begin{equation}
\theta = \pi+\arctan \frac{4\pi Q}{\beta A} - \arctan \frac{4\pi Q}{\gamma\eta M_s A}. \label{eq:Hall_angle}
\end{equation}
The Hall angle vanishes, if (C1) $Q=0$ or (C2) $\beta=\gamma\eta M_s$.
The value of velocity is proportional to the current:
\begin{equation}
|\dot {\mathbf R}| = \bigg(\frac{16\pi^2Q^2+\beta^2A^2}{16\pi^2Q^2+\gamma^2\eta^2 M_s^2A^2}\bigg)^{\frac12}j. \label{eq:module_velocity}
\end{equation}
In the case (C2) the velocity does not depend on the topological charge
and on the dissipation constant.
In the case $Q=0$, the soliton velocity is given by:
\begin{equation}
|\dot {\mathbf R}| = \frac{\beta}{\gamma\eta M_s}j,
\end{equation}
and is determined by ratio of the damping constants.

Thus, the Thiele equation predicts that the skyrmions localized in the AFM domains, will have no Hall angle due to the vanishing of the topological charge, and the sign of the Hall effect for the skyrmions in FM domains depends on the ratio of the Gilbert damping $\eta$ and STT torque $\beta$. We will further confirm these predictions in the numerical simulations.

For the case, when the skyrmion in one of the layers is in the vicinity of the domain wall in the other layer, one can also employ the Thiele equation. We assume rotational symmetry of skyrmion (in practice the symmetry can be slightly violated
due to interaction with the domain wall).
For clarity we consider flat grain boundary and straight domain wall, that is
the IEC potential $\Phi$ and the domain wall texture $\mathbf{n}_2$ depend only on $x$ coordinate.
Energy of the system up to an additive constant is given by:
\begin{equation}
V = -J_{1,2}\int \Phi(x) \mathbf{n}_1(x-R_x^1,y-R_y^1)\cdot \mathbf{n}_2(x-R_x^2)\,dx\,dy.
\end{equation}

The magnetization $M_s$, the current $j$ and damping parameters $\eta$, $\beta$ are assumed equal in both layers.
Thiele equation for the system is

\begin{equation}\label{eq:DW1}
\begin{cases}
    \gamma\eta M_s A\dot R_x^1 + 4\pi Q \dot R_y^1   = \dfrac{\gamma}{M_s}\dfrac{\partial V}{\partial R_x^1}-\beta Aj_x-4\pi Qj_y,\\
    -4\pi Q \dot R_x^1 + \gamma\eta M_s A\dot R_y^1 = 4\pi Qj_x-\beta Aj_y,\\
    \gamma\eta M_s b\dot R_x^2   = \dfrac{\gamma}{M_s}\dfrac{\partial V}{\partial R_x^2}-\beta bj_x,\\
\end{cases}
\end{equation}

First equation in (\ref{eq:DW1}) defines two competing forces, acting on a skyrmion on the grain boundary: the first one from potential gradient and the second one from current. 
Railing behavior of skyrmion is observed for the values of perpendicular current less than some critical value, when the returning force can no longer compensate the action of electric current and the skyrmion leaves the grain boundary.

\begin{equation}
   j_x^{cr} = \frac{\gamma^2\eta A}{ \Big(4\pi Q\Big)^2 + \Big(\beta A\Big)^2}\cdot\frac{\partial V}{\partial R_x^1} = \kappa\cdot\frac{\partial V}{\partial R_x^1}  
 \label{eq:Thiele:Domain_wall}
\end{equation}

\begin{figure}[!h]
\center{\includegraphics[width=0.45\textwidth]{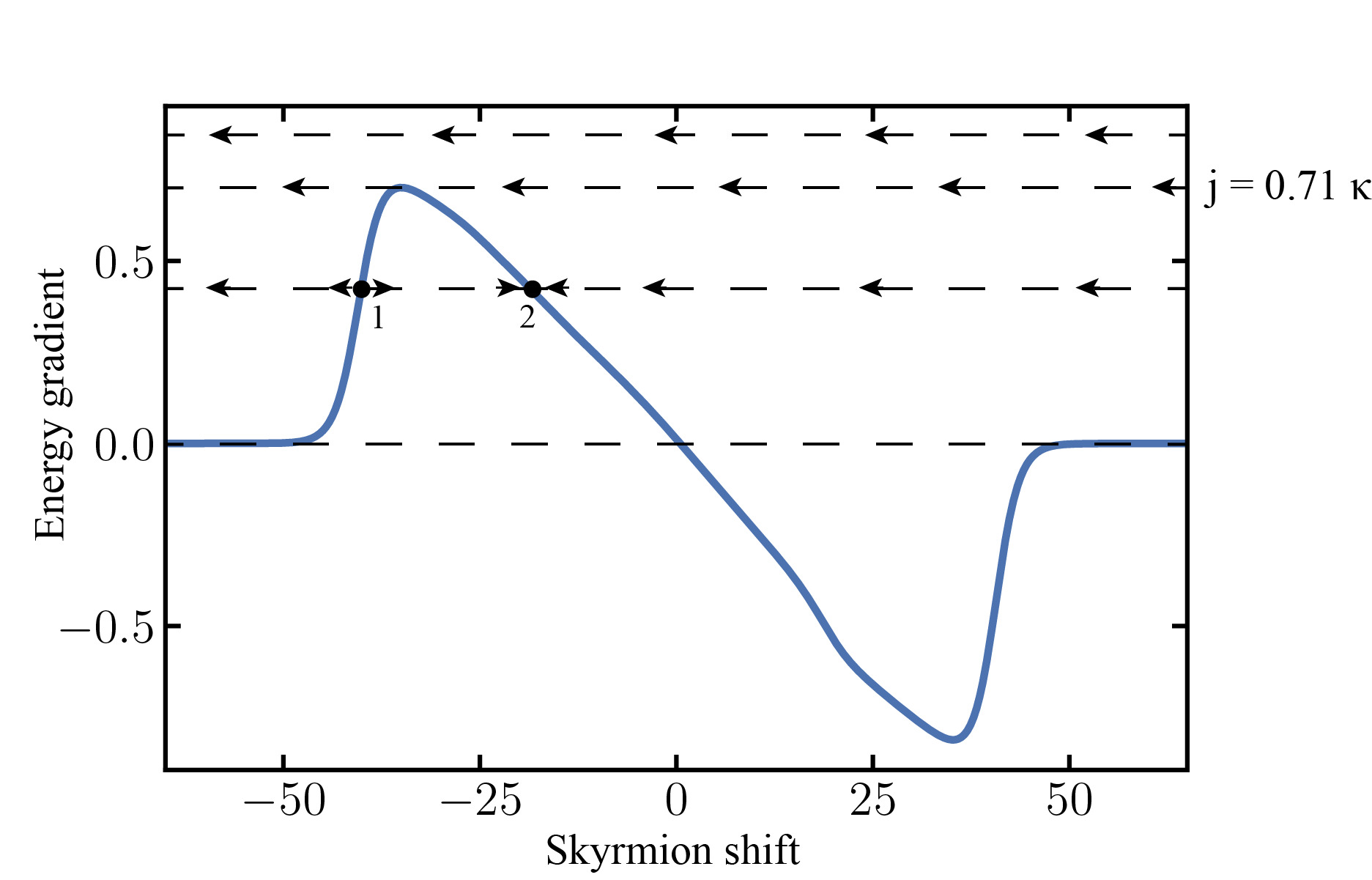}}
\begin{centering}
\caption{The blue line defines the energy gradient as a function of skyrmion shift from the equilibrium position, which is actually the returning force.  Energy profile, equilibrium
position and interaction potential are given on Figure 4. In order to simplify the analysis of skyrmion motion, we consider the special case of the dissipation constants when the Hall angle vanishes: $\beta=\gamma\eta M_s$. In this case the perpendicular force is produced only by $j_x$ and by the potential gradient.
Dashed horizontal lines demonstrate the absolute value of perpendicular current, given in the units of energy gradient.  Arrows define the direction of skyrmion motion: we see, that for the current larger, than critical value $0.71 \kappa$ skyrmion leaves the boundary of the grain. For smaller currents we obtain two equilibrium position, one of which is sustainable, and the other one is unsustainable.}
\label{fig:Fig5}
\end{centering}
\end{figure}

Here we introduced the parameter $\kappa$:

\begin{equation}
    \kappa= \frac{\gamma^2\eta A}{ \Big(4\pi Q\Big)^2 + \Big(\beta A\Big)^2}
\end{equation}
Fig.~\ref{fig:Fig5} demonstrates the equilibrium points and critical current for the case 
of pinning to the rail.

\begin{figure}[!h]
\center{\includegraphics[scale=0.21]{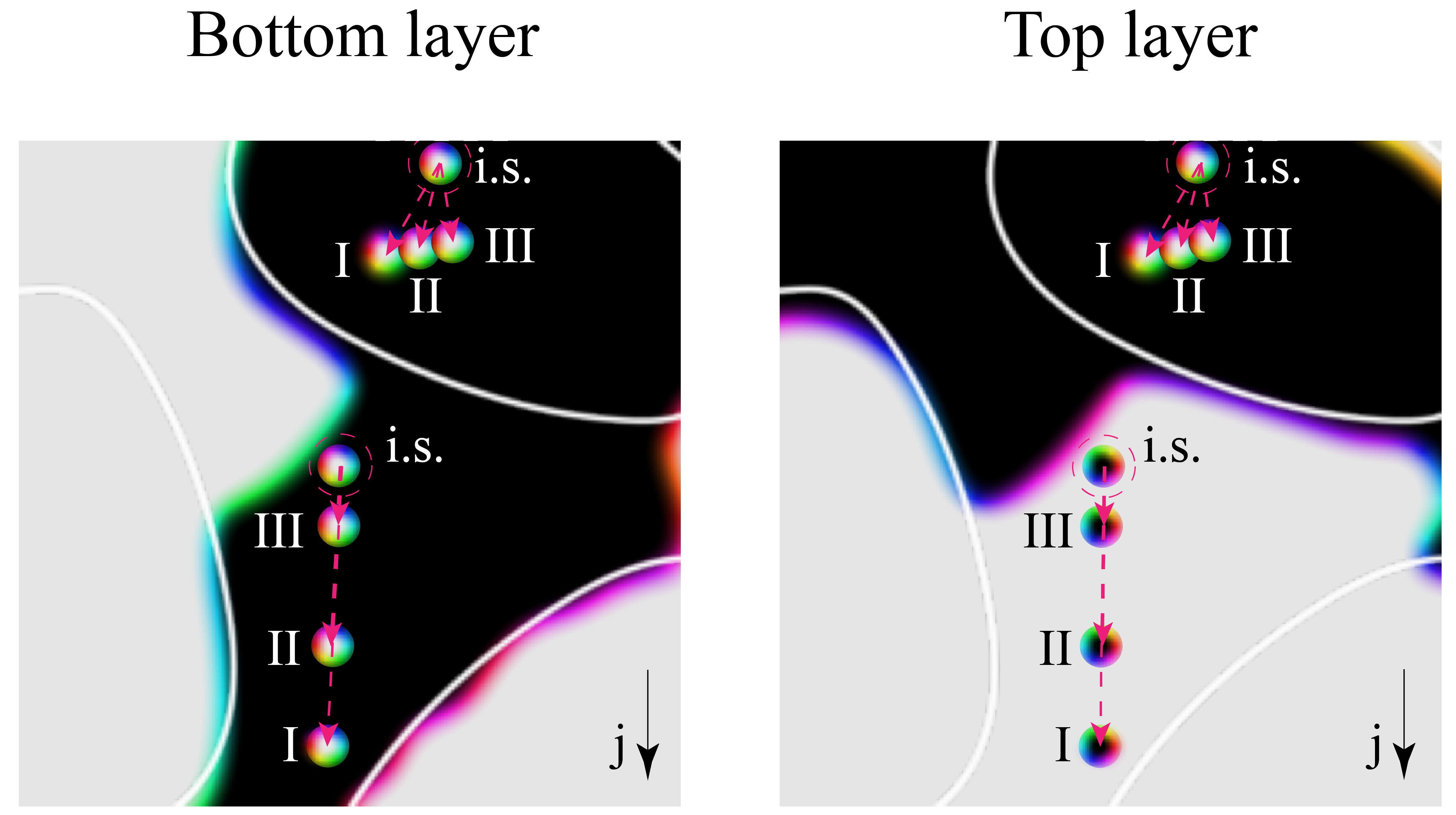}}
\begin{centering}
\caption{Dynamics of skyrmion pairs for different values of $\beta$. The temporal evolution of skyrmion positions over an identical period of time is demonstrated. i.s. - initial state, I - final state for $\beta$ = 0.6, II - $\beta$=0.45, III - $\beta$=0.15. The value of Gilbert damping constant $\eta$ here equals 0.2. The color scheme description is the same as in Fig. \ref{fig:Fig3}.}
\label{fig:Fig6}
\end{centering}
\end{figure}

We now check the predictions of the Thiele equation by numerical simulations via LLG equation. The results of the simulations for the FM and AFM skyrmion pairs are shown in Fig.~\ref{fig:Fig6}. The simulation was performed by evolving the initial state (labelled "i.s." in the figures) for a fixed amount of time and taking the snapshot of the final state. The simulation was performed for three different values of the antidamping constant $\beta$. First it is seen that, for the AFM skyrmion pair with zero topological charge, the Hall angle vanishes and the skyrmion propagates along the current according to Eq.~\ref{eq:Hall_angle}. Moreover, the module of velocity grows linearly with $\beta$ as follows from Eq.~\ref{eq:module_velocity}. For the case of FM skyrmion, the Hall angle is generally finite and depends on the ratio between antidamping constant $\beta$ and Gilbert damping constant $\eta$. 

\begin{figure}[!h]
\center{\includegraphics[scale=0.21]{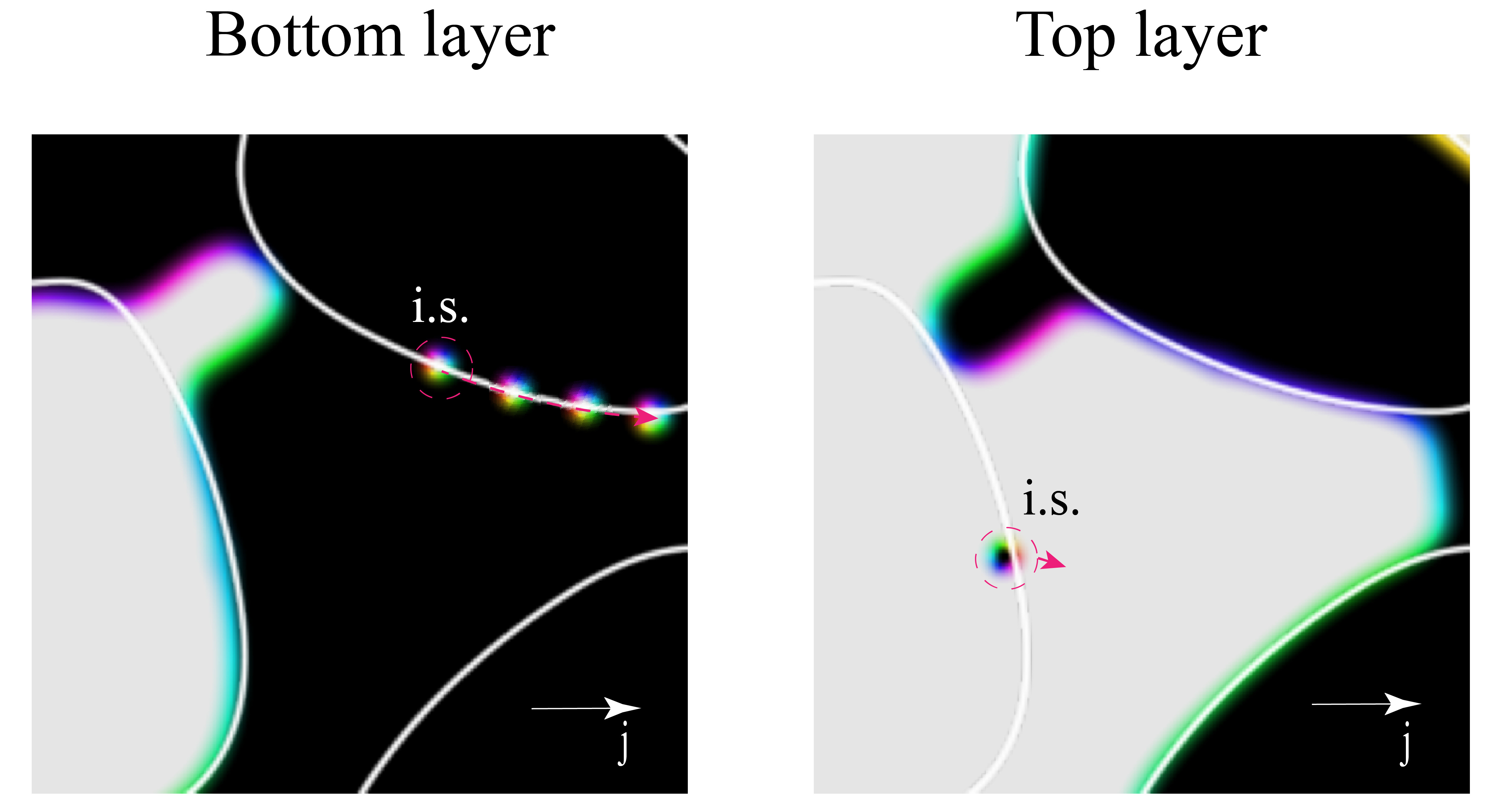}}
\begin{centering}
\caption{Motion of skyrmion along the domain wall. i.s. - initial state. The color scheme description is the same as in Fig. \ref{fig:Fig3}. Single skyrmions tend to be located in the area of zero interlayer interaction potential (borders of moir{\'e} grains). If current direction makes not too large angle with the direction of the grain boundary, the skyrmion remains attached to this boundary in the process of movement and moves along a kind of rail.}
\label{fig:Fig7}
\end{centering}
\end{figure}

For the single layer skyrmion localized in the vicinity of the domain wall in the other layer, the LLG simulations are shown in Fig.~\ref{fig:Fig7}. In the figure, two skyrmions can be identified localized in two two different layers. It can be seen, that for the bottom layer skyrmion, the angle between the current direction and the domain wall is relatively small and thus skryrmion propagates along the domain wall. At the same time, for the top layer skyrmion, the current is almost perpendicular to the domain wall and thus the skyrmion is dragged way from the domain wall and then disappears. This behaviour qualitatively corresponds to the predictions of the Thiele eq.~\ref{eq:Thiele:Domain_wall}. As can be seen, the Thiele equation gives qualitatively correct predictions for the dynamics of the three types of the skyrmions. This is due to the fact that as we show in Supplemental material, the skyrmion profile indeed remains almost unaffected in the course of motion under applied current.

\section{Conclusions}

We have shown that the interplay of spatially dependent interlayer moir{\'e} exchange potential and 
Dzyaloshinskii-Moriya interaction in van der Waals magnets leads to a rich variety of non-collinear magnetic structures. 
Specifically, we have identified three distinct families of skyrmions characterized 
by different topological properties and kinetics under applied in-plane current. 
Of particular interest are the skyrmions pinned to the grain boundary of the moir{\'e} potential. 
Our numerical calculations predict the railing of these skyrmions along the grain boundary under applied current and 
we have provided an analytical description of this effect using the Thiele equation. 
This behaviour is quite general for the two-layer structures with spatially varying interlayer exchange potential 
and we thus anticipate, that it may be observed in different vdW moir{\'e} magnetic bilayers. 
Railing of skyrmions in vdW magnets opens  routes towards novel applications of these heterostructures in spintronics.

\section{Acknowledgements}

The study was supported by the Russian Science Foundation 
grant N 22-22-00632, https://rscf.ru/project/22-22-00632/.

\appendix
\section{Non-collinear magnetic phases in twisted magnetic bilayer}

\begin{figure}[!h]
\center{\includegraphics[scale=0.15]{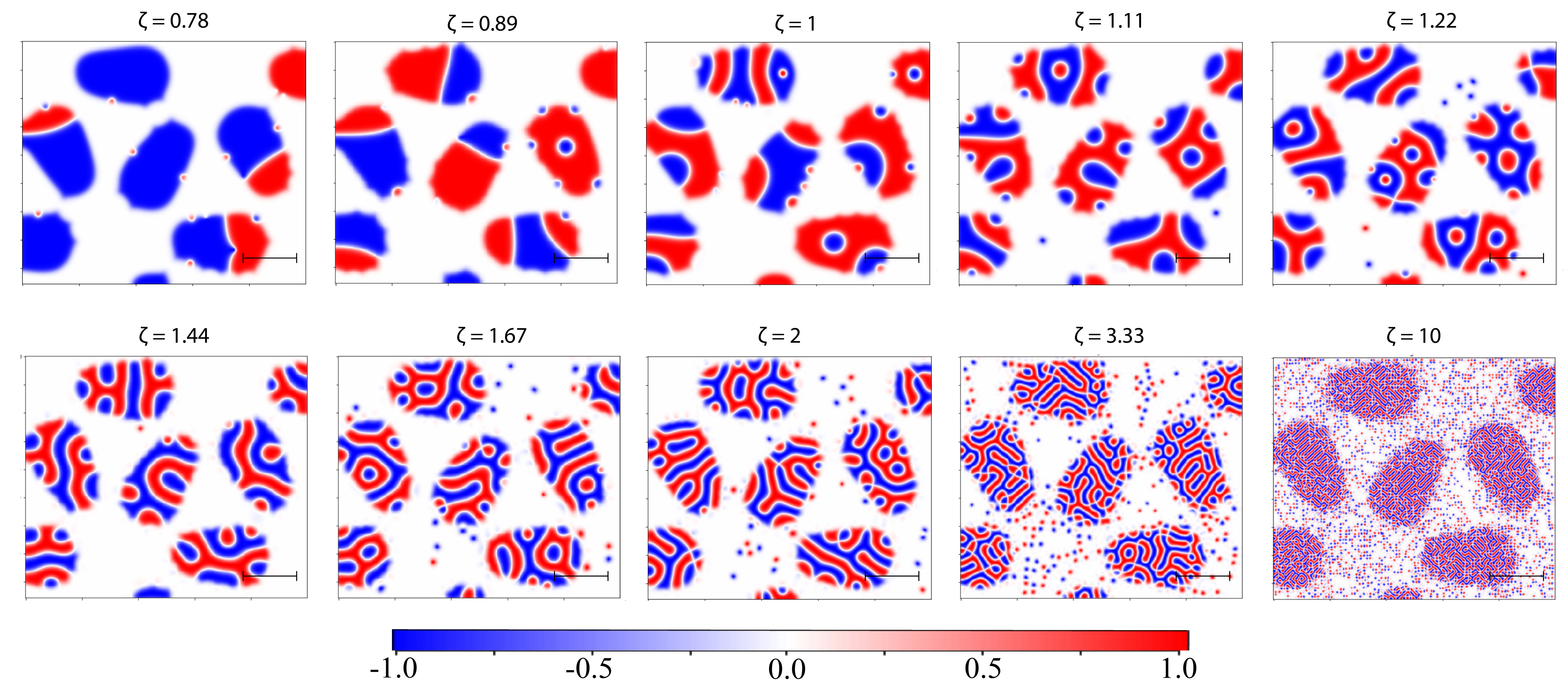}}
\caption{z-projection of total bilayer magnetic moments in moir{\'e} magnets for different values of $\zeta$.  Scale bar: 100 lattice constants.}
\label{fig:Fig1_sup}
\end{figure}

Fig.~\ref{fig:Fig1_sup} provides the evolution of non-collinear phases, stabilized in moir{\'e} magnets for different values of $\zeta$. The 
dimensionless parameter $\zeta=D/D_s$.

\begin{figure}[!h]
\center{\includegraphics[scale=0.6]{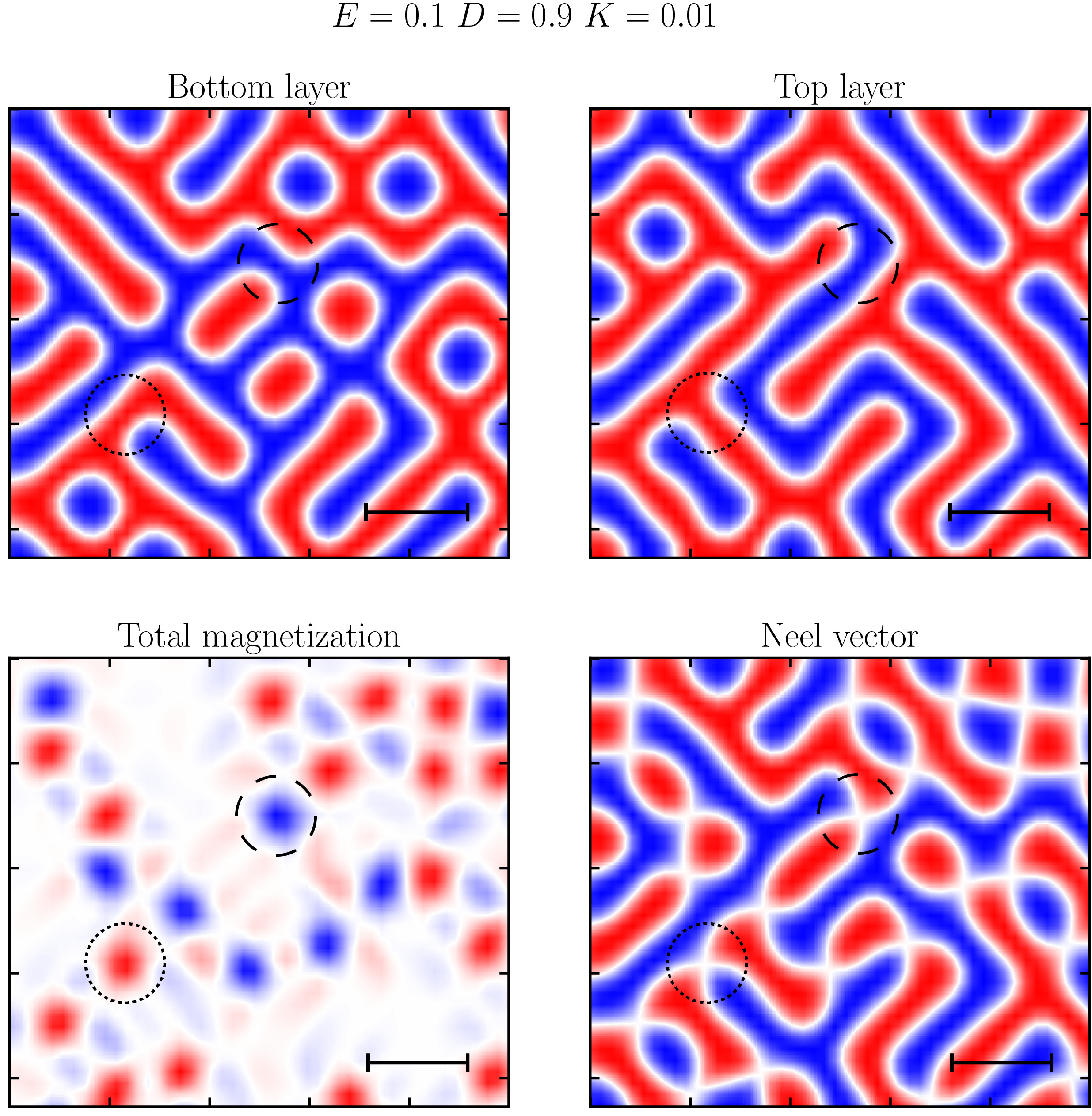}}
\caption{Normalized z-projection of magnetic moments for $\zeta=10$. Magnetization in the first layer, second layer, total magnetization (the sum of the magnetizations of the first and second layers at a given point), and Neel vector (the difference between the magnetizations of the first and second layers) are given. Scale bar: 10 lattice constants.}
\label{fig:Fig2_sup}
\end{figure}

More detailed consideration of magnetization for $\zeta$ = 10 is given on Fig.~\ref{fig:Fig2_sup}. The structure obtained in that case is the so-called "fingerprint" pattern. The small points that manifest in the regions characterized by antiferromagnetic exchange coupling are subjected to a detailed investigation. These defects emerge at the intersections of domains ("fingers"), as visually exemplified in the figure.  

\onecolumngrid

\twocolumngrid

\section{Thiele equation}

\begin{figure}[H]
\center{\includegraphics[scale=0.32]{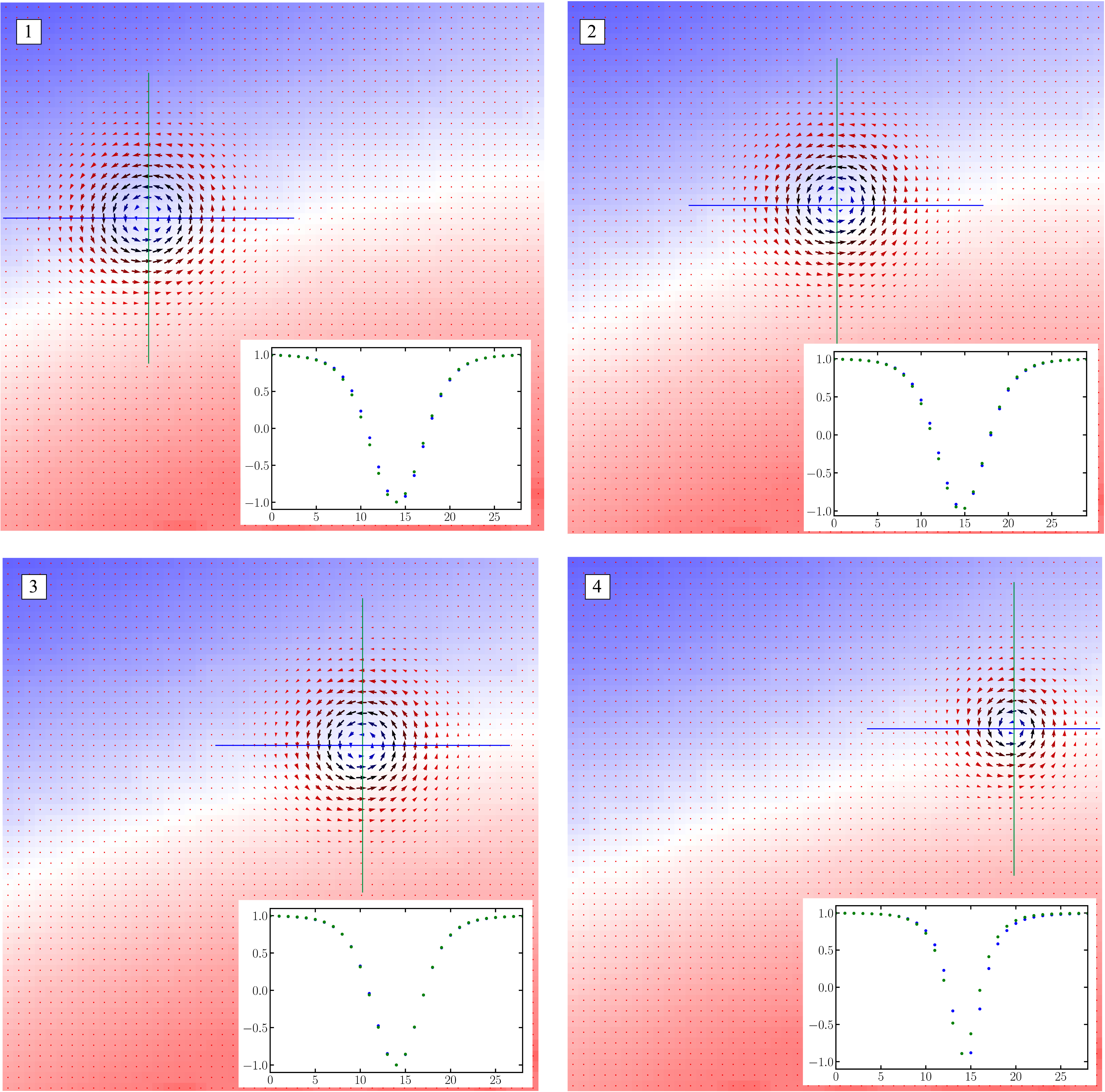}}
\caption{Modification of the skyrmion profile as it moves along the domain wall. Insets show two cross-sections of the $z$ component of the skyrmion magnetization along the lines of corresponding color}
\label{fig:Fig3_sup}
\end{figure}

To discover the applicability of the Thiele equation in our analysis, we examine the deformation of the skyrmion profile, as depicted in Fig.\ref{fig:Fig3_sup}. Notably, the skyrmion profile remains predominantly unchanged and retains its symmetry while moving along the domain wall. However, once the skyrmion starts leaving the boundary and begins to collapse, deformation becomes evident. Based on these observations, we infer that the Thiele equation holds validity during the stable motion of the skyrmion.

\section{Critical current condition}

Consider the system of equations, obtained for the case of skyrmion and domain wall interaction.

\begin{equation}
\label{eq:DW2}
\begin{cases}
\gamma\eta M_s A\dot R_x^1 + 4\pi Q \dot R_y^1   = \dfrac{\gamma}{M_s}\dfrac{\partial V}{\partial R_x^1}-\beta Aj_x-4\pi Qj_y,\\
    -4\pi Q \dot R_x^1 + \gamma\eta M_s A\dot R_y^1 = 4\pi Qj_x-\beta Aj_y,\\
    \gamma\eta M_s b\dot R_x^2   = \dfrac{\gamma}{M_s}\dfrac{\partial V}{\partial R_x^2}-\beta bj_x,\\
    \end{cases}
\end{equation}

Consider first two equations that relate to the first layer. We focuse on the special case, in which the Hall angle vanishes and the skyrmion tends to move along the current (this regime is provided by the special relation between the parameters: $\beta = \gamma\eta M_s$. Following this, the equations get the form:

\begin{equation}
\label{eq:DW3}
\begin{cases}
\beta A\dot R_x^1 + 4\pi Q \dot R_y^1   = \dfrac{\gamma}{M_s}\dfrac{\partial V}{\partial R_x^1}-\beta Aj_x - 4\pi Qj_y,\\
    -4\pi Q \dot R_x^1 + \beta A\dot R_y^1 = 4\pi Qj_x - \beta A j_y,\\
\end{cases}    
\end{equation}

\begin{equation}
\begin{cases}
    \dot R_y^1 = \dfrac{4\pi Q}{\beta A} \Big(j_x + \dot R_x^1\Big) - j_y,\\
    \dot R_x^1  = \dfrac{\gamma}{M_s\beta A}\cdot\dfrac{\partial V}{\partial R_x^1}-j_x - \dfrac{4\pi Q}{\beta A} \dot R_y^1 - \dfrac{4\pi Q}{\beta A} j_y,\\
\end{cases}
\end{equation}

\begin{equation}
\label{eq:DW4}
\dot R_x^1 = \frac{\gamma\beta A}{M_s \Big(\Big(4\pi Q\Big)^2 + \Big(\beta A\Big)^2\Big)}\cdot\frac{\partial V}{\partial R_x^1} - j_x  
\end{equation}

Condition for the critical current (the value of current when the returning force does not compensate the current any more and the skyrmion leaves the boundary): $\dot R_x^1 = 0$. Here we introduced the parameter $\kappa$. 

\begin{equation}
\label{eq:DW5}
   j_x^{cr} =  \dfrac{\gamma^2\eta A}{ \Big(4\pi Q\Big)^2 + \Big(\beta A\Big)^2}\cdot\frac{\partial V}{\partial R_x^1} = \kappa\cdot\frac{\partial V}{\partial R_x^1}  
\end{equation}

\onecolumngrid

\twocolumngrid

\bibliography{sample}

\end{document}